%% file: main.tex
\documentclass[sigconf,nonacm]{acmart}
\pdfoutput=1

\input preamble/preamble

\input preamble/macros

\keywords{User Tracking, GDPR, User Consent, Web fingerprinting}
\begin{document}
\title{User Tracking in the Post-cookie Era: How Websites Bypass GDPR Consent to Track Users\\
\subtitle{Please cite our WWW'2021 paper, doi: 10.1145/3442381.3450056}
}

\author{Emmanouil Papadogiannakis}
\affiliation{
	\institution{FORTH/University of Crete, Greece}
}
\author{Panagiotis Papadopoulos}
    \affiliation{
	\institution{Telefonica Research, Spain}
}
\author{Nicolas Kourtellis}
\affiliation{
	\institution{Telefonica Research, Spain}
}
\author{Evangelos P. Markatos}
\affiliation{
	\institution{FORTH/University of Crete, Greece}
}

\newcommand*{\origrightarrow}{}
\let\oldarrow\textrightarrow
\renewcommand*{\textrightarrow}{\fontfamily{cmr}\selectfont\origrightarrow}

\input sections/00_abstract
\maketitle
\pagestyle{plain}

\input sections/01_introduction
\input sections/02_background
\input sections/03_crawling
\input sections/04_measurements
\input sections/05_edge_observations
\input sections/06_related
\input sections/07_discussion
\input sections/08_conclusion
\input sections/09_acknowledgments

\bibliographystyle{unsrt}
\balance
\bibliography{main}
\end{document}

%% file: preamble/preamble.tex
\usepackage{booktabs}
\usepackage{pbox}
\usepackage{epsfig,endnotes}
\usepackage{epstopdf}
\usepackage{xcolor}
\usepackage{graphicx}
\usepackage{xspace}
\usepackage{enumitem}
\usepackage{url}
\usepackage{balance}
\usepackage{caption}
\usepackage{subcaption}
\usepackage{breakurl}
\usepackage{pbox}
\usepackage{times}  
\usepackage{pdfrender}
\usepackage{csquotes}

%% file: preamble/macros.tex
\newcommand{\eg}{e.g.,\xspace}
\newcommand{\etc}{etc.\xspace}
\newcommand{\etal}{et al.\xspace}
\newcommand{\ie}{i.e.,\xspace}
\newcommand{\noaction}{\texttt{No Action}\xspace}
\newcommand{\acceptall}{\texttt{Accept All}\xspace}
\newcommand{\rejectall}{\texttt{Reject All}\xspace}
\newcommand{\sitesCrawled}{850K\xspace}
\newcommand{\sitesResponded}{628,213\xspace}
\newcommand{\sitesCMP}{27,953\xspace}
\newcommand{\sitesCMPNoError}{27,180\xspace}

\newcommand{\point}[1]{\vspace{.05in} \par\noindent\textbf{#1}:\xspace}

\newcommand{\idleaking}{first-party ID leaking\xspace}
\newcommand{\Idleaking}{First-party ID leaking\xspace}
\newcommand{\idleaks}{first-party ID leaks\xspace}

\newcommand{\percentileLabels}{min, 25th percentile, median, 75th percentile, max\xspace}
\newcommand{\thirdpartiesnoactionMedian}{16\xspace}
\newcommand{\thirdpartiesrejectallMedian}{17\xspace}
\newcommand{\thirdpartiesacceptallMedian}{19\xspace}

\newcommand{\idleakingnoactionAverage}{2.14\xspace}
\newcommand{\idleakingrejectallAverage}{2.49\xspace}
\newcommand{\idleakingacceptallAverage}{3.04\xspace}

\newcommand{\csyncnoactionAverage}{3.51\xspace}
\newcommand{\csyncrejectallAverage}{3.91\xspace}
\newcommand{\csyncacceptallAverage}{4.86\xspace}

%% file: sections/00_abstract.tex
\begin{abstract}

During the past few years, mostly as a result of the GDPR and the CCPA, websites have started to present users with cookie consent banners.
These banners are web forms where the users can state their preference and declare which cookies they would like to accept, if such option exists.
Although requesting consent before storing any identifiable information is a good start towards respecting the user privacy, yet previous research has shown that websites do not always respect user choices.
Furthermore, considering the ever decreasing reliance of trackers on cookies and actions browser vendors take by blocking or restricting third-party cookies, we anticipate a world where stateless tracking emerges, either because trackers or websites do not use cookies, or because users simply refuse to accept any.
In this paper, we explore whether websites use more persistent and sophisticated forms of tracking in order to track users who said they do not want cookies.
Such forms of tracking include \idleaking, ID synchronization, and browser fingerprinting.
Our results suggest that websites do use such modern forms of tracking even before users had the opportunity to register their choice with respect to cookies.
To add insult to injury, when users choose to raise their voice and reject all cookies, user tracking only intensifies.
As a result, users' choices play very little role with respect to tracking: we measured that more than 75\% of tracking activities happened before users had the opportunity to make a selection in the cookie consent banner, or when users chose to reject all cookies. 

\end{abstract}

%% file: sections/01_introduction.tex
\section{Introduction}
\label{sec:intro}

Over the past few years, we have seen an increasing concern about user data protection with respect to the data of European users.
This was probably the result of the General Data Protection Regulation (GDPR) which was adopted in April 2016 and came into force in May 2018.
The main difference of this regulation compared to previous legislation is that it includes significant fines for companies which collect users data without the users' consent or some other legal basis.
Such fines can reach up to 20 million euros, or up to 4\% of the annual worldwide turnover of the preceding financial year, whichever is greater.
As a result, several companies, and their associated websites, have started asking their visitors and users for their consent, before collecting (and processing) their data.

Such a consent has been usually collected via cookie banners, which ask users for consent and may give some choices as well.
Indeed, users may be given the choice to accept all cookies, to accept some cookies, or even to reject all cookies.
The choice is entirely up to the user, and the correct implementation of this choice is the responsibility of the website.
Although this sounds completely legal and fully straightforward, deviations have been reported in literature~\cite{fouad:hal-02567022,eijk2019impact,10.1145/3321705.3329806,9152617,utz2019informed}.
For example, some websites claim that some cookies are absolutely necessary for their operation (\eg~for the page to be delivered) or due to legitimate interest (\eg~to improve the product), and can not be rejected by the users.
Thus, users cannot really choose to reject \emph{all} cookies: these necessary cookies cannot be rejected.
Past research studies have noticed some discrepancies between what the users type and what is registered in the website.
For example, the users may provide a negative response (\ie reject all cookies), but the cookie banners may register a positive one (\ie accept all cookies), or the cookie banners may register a positive response even before the users had the opportunity to provide any choice~\cite{9152617}.

All these previous studies focus on cookies and compliance of cookie processing with the GDPR.
In this paper, we set out to explore a slightly different question:
\begin{quote}
\emph{
    If a user does not provide consent, or chooses to \textbf{reject} all cookies, do websites use other forms of tracking to track this user?
    If so, what are these forms of tracking, and what is the extent of this tracking?
    }
\end{quote}

Considering the (i) ever less reliance of third-party trackers on non-permanent, erasable state-like cookies~\cite{mervis2020cookieless} and (ii) recent advances of browser vendors against third-party cookies~\cite{cookiebot2021chrome-cookies,wilander2019tracking-prevention}, it is apparent that the need for identifying how websites treat user consent in case of stateless (cookie-less) tracking is more than timely and urgent.
We address this need and try to fill this exact gap in our understanding, by being the first to investigate what is the GDPR compliance across the Web in the case where websites and trackers do not use cookies, or users do not accept cookies.

Sadly, our results suggest that even when users reject all cookies, websites \emph{do use other forms of tracking} to track users, and process personal data, in violation of GDPR.
Such forms of tracking include \emph{\idleaking}, \emph{ID synchronization} and \emph{browser fingerprinting}.\footnote{One might think that ID synchronization is a form of tracking using cookies.
This is not really true: although ID synchronization does use (values stored in) cookies, passing such values around is done in an unorthodox manner, completely different from the way cookies are used.}
\Idleaking and ID synchronization are used to pass an identifier (such as a cookie) as an ``argument'' in an HTTP request to a website - different from the website that planted this ID in the first place.
In fact, according to past studies~\cite{papadopoulos2019cookie,10.1145/3178876.3186060,falahrastegar2016tracking}, Web entities may share IDs they have assigned to users and help third-parties re-identify users or create universal IDs.
Browser fingerprinting~\cite{englehardt2016online,acar2014web} uses elaborate approaches to uniquely identify a user through characteristics of her device - characteristics which can be easily found by a website.
Such characteristics may include screen resolution and rendering characteristics, browser fonts and installed plugins \etc~\cite{eckersley2010unique,mayer2012third,nikiforakis2013cookieless,papadopoulos2017long}.
Combining several of these characteristics can provide a large enough number of entropy bits to uniquely identify a user.

Although these cases of user identification are considered ``personal data processing'' according to GDPR and ePrivacy~\cite{eprivacy2017} regulations, and must be visible to users, they often do not appear in request forms of consent managers deployed by modern websites.
In this study, we highlight exactly that: the lack of transparency and user consent when it comes to websites that deploy user identification techniques like ID synchronization and browser fingerprinting.

\noindent 
The contributions of this work are as follows:
\begin{itemize}[leftmargin=0.5cm]

    \item We propose a fully automated method for detecting browser fingerprinting on websites using the Chromium Profiler.
    
    \item We crawl close to one million websites and record how they track users using sophisticated forms of tracking (such as \idleaking, ID synchronization and browser fingerprinting) as a function of users' choices.
    
    \item We find that: (1) More than 75\% of leaks happen despite the fact that users have chosen to reject all cookies; (2) Websites embedded with ID synchronizing third-parties force browsers to engage in several ID synchronizations (\csyncnoactionAverage per ID, on average) even before users had a chance to accept or deny consent; (3) Less popular websites are more likely to disregard users’ consent choices and engage in \idleaking  and ID synchronization; (4) Browsers leak more information when users choose to reject all cookies than when they choose to take no action at all; (5) Our analysis of tracking per country code reveals significant discrepancies across EU countries.
    
    \item Our methodology can be transformed into an auditing tool for regulators, stakeholders and privacy-policy makers, for verifying compliance with GDPR and users' privacy rights.
    
    \item We make our crawling and analysis tool publicly available~\footnote{https://gitlab.com/papamano/consent-guard} to support further research on this topic.
    
\end{itemize}

%% file: sections/02_background.tex
\section{Background}
\label{sec:background}

In the world of Web, cookies are used to store identifying information for a given user.
However, recent policies and regulations from browser vendors and government bodies~\cite{sameorigin,itp,cookielaw} try to control the exposure of this identifying information to third-parties and for how long.
These policies restrain the ad and tracking industry that relies on re-identifying a user for long periods to serve more targeted ads.
Some of the most popular techniques used by the third-parties include ID synchronization (\eg cookie synchronization~\cite{papadopoulos2019cookie,acar2014web,10.1145/3193111.3193117,olejnik2013selling}) and canvas fingerprinting~\cite{mowery2012pixel}, but also the font-based fingerprinting~\cite{eckersley2010unique}, WebRTC-based fingerprinting, AudioContext fingerprinting, and Battery API fingerprinting~\cite{englehardt2016online}.

\subsection{ID Sharing}
\label{sec:csync}

\begin{figure}[t]
    \centering
    \includegraphics[width=0.7\columnwidth]{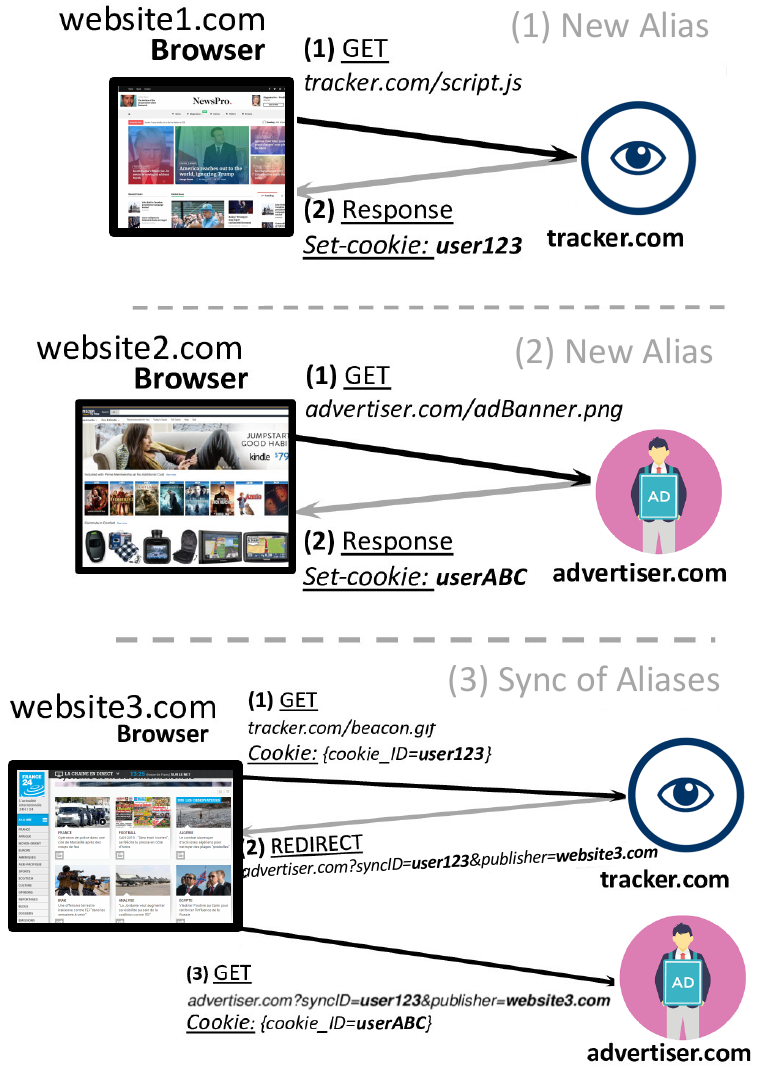}
    \vspace{-0.3cm}
    \caption{Example of an ID synchronization operation. Two entities match the IDs they have assigned to the same user.}
    \label{fig:csync}
\end{figure} 

Whenever a user visits a new website, a plethora of cookies and IDs are assigned to her, allowing first or third-parties to re-identify her across the Web and build a profile based on her browsing behavior.
These profiles can be later centralized in Data Management Platforms~\cite{10.14778/2536222.2536238}, sold by data brokers~\cite{databrokers}, or used by advertisers to bid in ad auctions~\cite{pachilakis2019no}, ad-retargeting~\cite{iordanou2019eyewnder} and cross-device tracking~\cite{solomos2019cdt-raid}.
For the different Web entities (\eg publishers, analytics, data brokers, advertisers, \etc) to perform such transactions, all of the different assigned aliases (\ie IDs) that each entity has assigned to the same user, need to be linked (\ie synced) together.
This would reveal that the user that the entity \texttt{A} knows as \texttt{userABC} is the same user that entity \texttt{B} knows as \texttt{user123}.

Figure~\ref{fig:csync} illustrates an example of how this ID synchronization takes place.
Assume a user browsing \texttt{website1.com} and \texttt{website2.com}, in which there are third-parties like \texttt{tracker.com} and \texttt{advertiser.com}, respectively.
Consequently, these two third-parties have the chance to assign an alias to the user and re-identify them in the future.
From now on, \texttt{tracker.com} knows the user with the ID \texttt{user123}, and \texttt{advertiser.com} knows the same user with the ID \texttt{userABC}.
Next, assume that the user lands on \texttt{website3.com}, which includes some JavaScript code from \texttt{tracker.com} making the browser issue a GET request to \texttt{tracker.com} (step 1), who responds back with a REDIRECT request (step 2), instructing the user’s browser to issue another request to its collaborator \texttt{advertiser.com} this time, using a specifically crafted URL (step 3) where the alias it uses (\ie \texttt{user123}) is piggybacked.
When \texttt{advertiser.com} receives the above request from the user it knows as  \texttt{userABC}, it learns that the user whom \texttt{tracker.com} knows as \texttt{user123}, and the user \texttt{userABC} are basically the same user.
This allows the two entities to join the different aliases (\eg cookies, device IDs, user IDs, \etc) a user has on the Web.

In this paper, we study two types of ID sharing: (i) \emph{\idleaking}, where a first-party alias (\eg a cookie or device ID) is leaked from the visited website to different third-parties, and (ii) \emph{third-party ID synchronization}, where third-parties link together the different third-party aliases they use for the same users.

\subsection{Browser Fingerprinting}
\label{sec:canvas}

\begin{figure}[t]
    \centering
    \includegraphics[width=0.9\columnwidth]{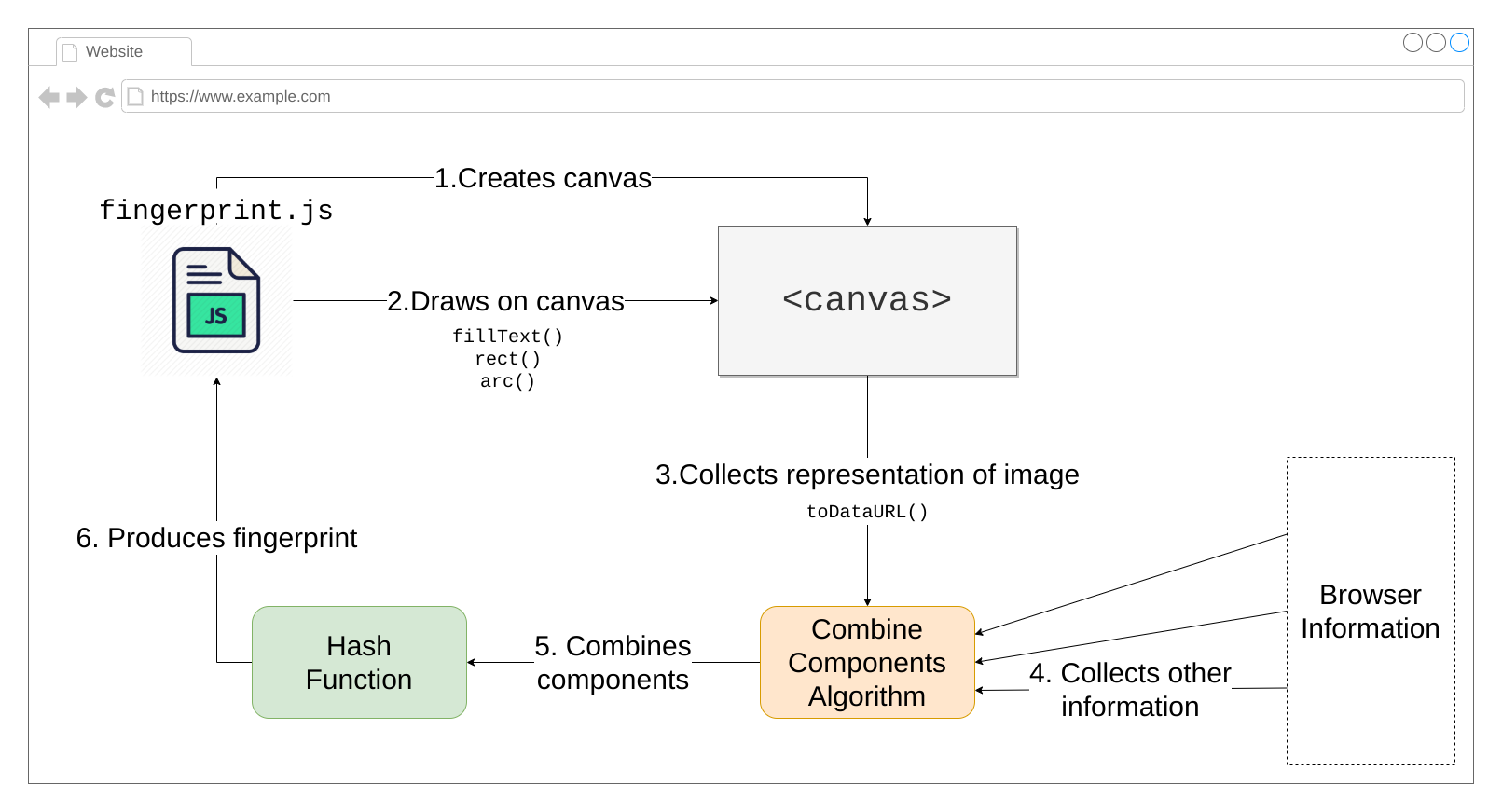}
    \vspace{-0.3cm}
    \caption{Canvas Fingerprinting process as part of the browser fingerprinting methodology used by popular libraries. The website can extract a fingerprint of the user's browser.}
    \vspace{-0.3cm}
    \label{fig:canvasFingerprintint}
\end{figure} 

Browser Fingerprinting is a sophisticated set of techniques, which can be used to uniquely identify browser instances without storing any information on the user side (stateless).
It can be used to detect malicious users that create multiple accounts in social networking services, or even stop deceitful orders in e-commerce platforms.
However, this technique can be abused by privacy-violating websites and, therefore, track users across sites, or even de-anonymize private sessions.
In fact, previous work~\cite{mowery2012pixel,laperdrix2016beauty} has shown that this technique provides sufficient bits of entropy to effectively track users, even through the usage of the Tor Browser.

One of the most prevalent and stealthy such fingerprinting techniques is Canvas Fingerprinting: named after the HTML canvas element, which was introduced in the latest version (\ie HTML5).
A canvas element provides the required functionality for drawing graphics using client-side code.
Moreover, canvas fingerprinting relies on WebGL, a cross-platform JavaScript API that enables developers to render advanced graphics using shaders.
As a result, developers have access to rendering functionality, which is performed in a GPU, however, in an HTML context via the canvas element.

Figure~\ref{fig:canvasFingerprintint} demonstrates the process of canvas fingerprinting as part of browser fingerprinting.
Assume (i) a website that contains the fingerprinting code and (ii) a browser instance that can execute JavaScript code.
As a first step, the fingerprinting script creates a canvas element using the built-in interface provided by almost all modern browsers.
Next, the script renders some 2D graphics and text on the canvas.
Usually, the text that is drawn is a pangram.
This means, that it contains all the letters of the English alphabet in order to increase the number of entropy bits.
Different font sizes and font families result in a slightly different text that can affect the final fingerprint.
As a next step, the fingerprinting script needs to extract the content of the canvas and inspect its pixel values (step 3).
This is achieved using the method \texttt{toDataURL()}, provided by the canvas object.
This method returns the Base64 encoding of the canvas' content.
Based on various factors, including fonts that are installed on the user's machine, version of OpenGL and browser's rendering engine, this string can be sufficiently different per user.

Then, the script combines this canvas fingerprint with other information, which can be used as an additional source of entropy (step 4).
This information includes, among others, the host operating system and timezone, its screen resolution, installed plugins, preferred language set in the browser and number of logical processors available on the host.
The output of the combination algorithm is a long string that uniquely identifies the specific browser instance (step 5).
Finally, the identifier is hashed, to produce a fingerprint for this specific browser (step 6) and is usually sent across the network, or even stored as a cookie.

\begin{figure*}[t]
    \centering
    \includegraphics[width=1.7\columnwidth]{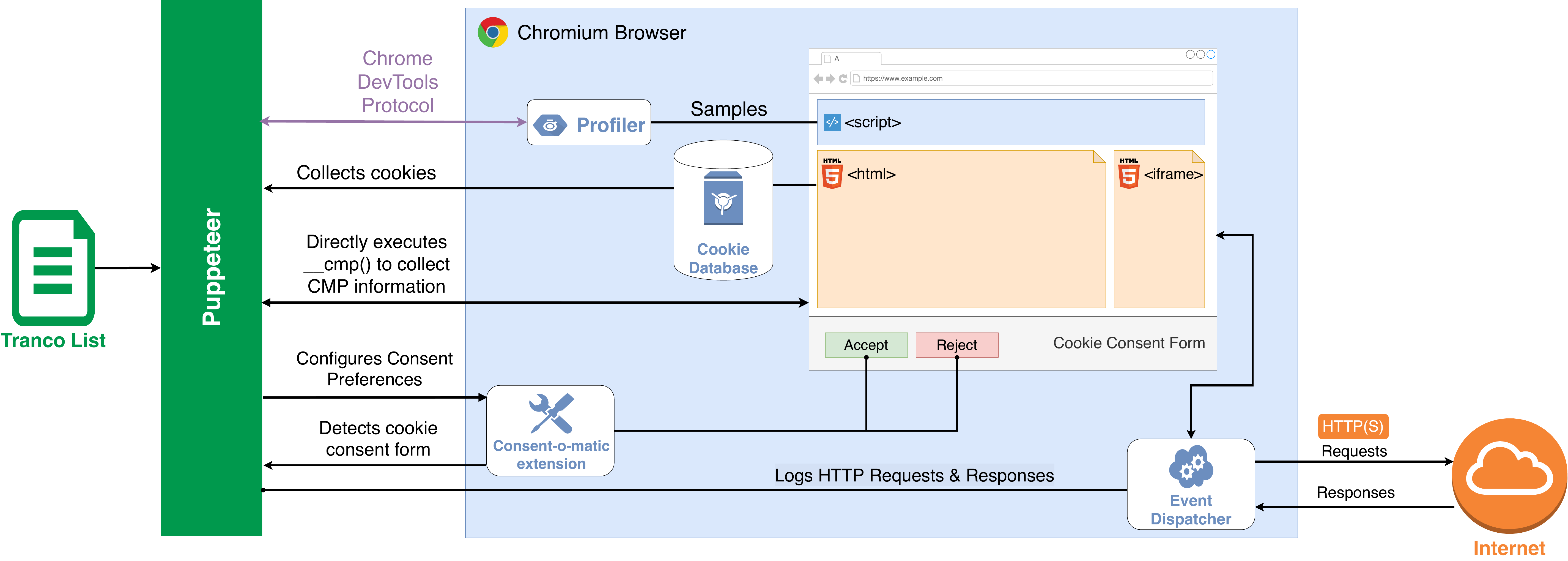}\vspace{-0.3cm}
    \caption{High level overview of our crawling methodology. We use Puppeteer to instrument a web browser and automatically visit websites. The Chrome Profiler is a built-in tool used to record and analyze run-time performance by collecting callsite information and execution statistics. The Cookie Database stores all cookies set by various domains. The Consent-O-matic tool is loaded on browser startup as an extension to handle cookie consent forms. Whenever a request is issued or a response is received, the event dispatcher emits the appropriate event, which is handled by our puppeteer-based crawler.}
    \label{fig:CrawlerArchitecture}
\end{figure*} 

Tracking techniques need to be transparent to users to avoid raising suspicion or harm the user experience.
As such, browser fingerprinting can be performed in minimum time on any browser that supports JavaScript by using invisible HTML elements and without requiring any privileges or permission from the user.
Consequently, even privacy-aware advanced users that block cookies can be tracked.
Furthermore, browser fingerprinting is difficult to prevent because it relies on native functionality, built in modern browsers.
Users need to either disable JavaScript, or use external browser extensions.
These techniques usually add random noise to some built-in functions, making the fingerprint different, each time the same website attempts to (re)identify a user~\cite{laperdrix2017fprandom, nikiforakis2015privaricator, brave2020}.

%% file: sections/03_crawling.tex
\section{Methodology}
\label{sec:methodology}

To investigate the effect of the different options a user is provided with while visiting websites with a consent form, we leverage the Consent-O-matic tool~\cite{consentomatic}.
Consent-O-matic is the state-of-the-art browser extension to automatically detect and handle GDPR consent forms.
Whenever the extension detects a Consent Management Platform (CMP), it logs its info (\eg vendor, encoding, IDs).
Additionally, it can be configured to either accept or reject the different categories of data processing purposes.
In addition to this, we develop a puppeteer-based crawler that instruments a Chrome browser.
By using Consent-O-matic, the browser can automatically perform one of the following three actions when a consent form is detected:
\begin{enumerate}
    \item \acceptall: grant consent for all data processing purposes to all third-parties residing in the visited website.
    \item \rejectall: deny consent for all data processing purposes to all third-parties residing in the visited website.
    \item \noaction: avoid interacting with the form in any way.
\end{enumerate}

By using our instrumented browser, we crawl (with clean state) the landing page of the top \sitesCrawled sites of Tranco list~\cite{tranco}.
This list aggregates the ranks from the lists provided by Alexa, Umbrella, and Majestic from 29.07.2020 to 27.08.2020 (pay-level domains retained)\footnote{https://tranco-list.eu/list/Q274/full}.
Whenever a CMP is detected, we crawl the given website 3 times (one for each of the different consent actions), and we store: the HTML, cookiejar, HTTP requests, HTTP responses, JS function calls and CMP info for each case.
It is important to note that, we capture HTTP(S) requests and responses passively, via the emitted Chrome events without mutating or intercepting them.
This ensures that the behavior of the website is not affected by our crawler.
An overview of our crawling methodology is illustrated in Figure~\ref{fig:CrawlerArchitecture}.

\begin{table}[t] 
\footnotesize
\caption{Summary of our crawled dataset.
}\vspace{-0.3cm}
    \centering
    \begin{tabular}{lrr}
    \toprule
        \textbf{Description}    & \textbf{Volume}   & \textbf{\% of total} \\ 
    \midrule
        Initial set of websites & \sitesCrawled             & \\ 
        Websites that errored   & 219,098                   & 25.78\% \\
        Websites that were filtered out & \\
        (pornographic or no-bots allowed) & 2,689           & 0.32\% \\
        Total websites correctly parsed & \sitesResponded   & 73.90\% \\
        Websites with a CMP             & \sitesCMP         & 3.29\% \\
        Websites with a CMP and no error & \\
        in all three consent actions & \sitesCMPNoError  & 3.20\% \\
        \bottomrule
    \end{tabular}
    \label{tab:dataset}\vspace{-0.3cm}
\end{table}

\subsection{Data Description and Analysis}
Overall, the crawler (located in EU) visited \sitesCrawled sites from August 28\textsuperscript{th} 2020 to September 17\textsuperscript{th} 2020, the Consent-O-matic extension detected \sitesCMP sites with a CMP (or 4.44\% of the successfully visited sites)\footnote{Inline with related works which report detection rates of 3\% ~\cite{10.1145/3321705.3329806} and 6.2\% \cite{9152617}.
Designing a detection tool with better accuracy is very challenging due to the heterogeneity of the various existing consent management libraries and custom solutions.}, and we collected a total of 108 GB of data for these sites.
Crawls failed at 25.78\% of the initial set of websites (due to error, puppeteer time-out, site inaccessibility, site did not serve EU-based users).
Table~\ref{tab:dataset} summarizes our dataset.

\point{Detecting Third-party ID Synchronization}
We perform an offline analysis on the collected data to detect ID synchronization operations.
Specifically, we examine all application-level network traffic and search for requests that contain unique IDs. For HTTP GET requests, we inspect the URL of the requests and examine their path and parameters.
For HTTP POST requests, we inspect the data stored in the request body.
We report a case of ID synchronization only if a unique ID is delivered to a domain different from the one that assigned it to the user.
This analysis is performed for both first-party and third-party set IDs and in a per-website base.
The majority of these IDs are stored in cookies.
Thus, we parse the value of each and look for strings that can be used as unique IDs.
If this value is a text string representing a JSON object, we get the values stored in key-value pairs in the object\footnote{We purposely ignore the keys found in key-value pairs of JSON objects, since these keys rely on the API which the website uses, and do not contain any useful information that can uniquely identify users.
Treating these keys as possible identifiers would result in multiple false positives.}.
If the object contains inner JSON objects, we recursively obtain all values in all nested levels.

To reduce false positives, we deliberately filter values that include consent information (\eg values of the keys \texttt{euconsent}, \texttt{eupubconsent}, \texttt{\_\_cmpconsent} and \texttt{\_\_cmpiab}).
As described in~\cite{9152617}, such values can be used to share user's consent across different CMPs or third-parties present on the page.
Additionally, we filter out values that are considered common and cannot be used as identifiers: strings that represent dates, timestamps, regions, locale, strings that end with a common file extension (\eg jpg), strings that are URLs (\eg start with www. or http://) and, finally, strings that are prevalent keywords.
To construct a list of such keywords, we use a simplified puppeteer-based crawler to visit over 2.5K websites, and store all cookies.
We manually inspect their values and we identify over 80 keywords that are frequently found in cookies but cannot be used for user identification.
This list includes keywords such as ``homepage'', ``undefined'', ``desktop'', ``not set'' and ``active'', among others.
We also exclude strings that have a length of 5 or less characters as they do not contain enough bits of entropy to uniquely identify a user.
In addition, we see cookie values combining (with a delimiter) identifiers with non-identifying info (\eg,  timestamp, locale, \etc), for example:  \emph{foo=\{userID\};15693242;en-US}.
We find less than 0.6\% of such IDs being synced with third-parties.

The last step is to detect the possible IDs in the HTTP traffic.
For each string of the previous step, we examine all HTTP requests targeting domains different than the one that set the cookie, and seek for an exact string match.
We search for these possible IDs in (i) URL parameters, (ii) the body of requests and (iii) the referrer header.
We tokenize the URL parameters using both default (\ie \&) and custom (\ie ``;'') delimiters.

\point{Detecting Browser Fingerprinting}
\label{sec:canvas_det}
As described in \cite{mowery2012pixel} and illustrated in Figure~\ref{fig:canvasFingerprintint}, browser fingerprinting techniques, such as canvas fingerprinting, can be performed using various native methods provided by the browser's run-time environment (\eg \texttt{fillText}).
Past work~\cite{acar2014web,raschke2018uncovering,englehardt2016online,le2017towards} has focused on monitoring these native methods along with their returned values.
By observing the sequence of function calls along with the arguments given to these functions, one can have indications of browser fingerprinting.
Additionally, searching for common arguments found in popular fingerprinting libraries can help increase the level of certainty.
We argue that this method produces multiple false positives, since websites which use the native methods or HTML elements, like the canvas element, legitimately, might be marked as fingerprinting websites.
Indeed, in~\cite{raschke2018uncovering} manual revision of results was required in order to exclude false positives.
To mitigate this, our approach focuses on a higher level of abstraction and does not examine the native (\ie browser's built-in) methods.
This way, we successfully disregard websites that use these methods legitimately (\eg the canvas element for web graphics).
Specifically, to detect browser fingerprinting, we perform JavaScript code profiling and search for specific function calls that indicate the presence of a fingerprinting library.
Our method reduces the number of true positives, but ensures that the results are trustworthy.
Moreover, this method can be utilized by a fully-automated crawler, without the need of manual intervention.
In particular, we analyse the open-source version of one of the most widely-used fingerprinting JavaScript libraries: FingerprintJS \cite{fingerprintjs}.
We extract the full list of functions used during the process of browser fingerprinting.
We then focus on functions that consist of multiple operations and require a significant number of execution cycles.
This ensures that they will always be sampled by the profiler.
Moreover, we ignore functions that have common names (\eg \texttt{map} or \texttt{isIE}) and functions that can be utilized by general purpose code to perform actions not necessarily related to fingerprinting (\eg \texttt{getRegularPlugins}).
As a result, we conclude that the execution of the functions \texttt{getCanvasFp}, \texttt{getWebglFp}, \texttt{Fingerprint2} and \texttt{Fingerprint2.get} indicate browser fingerprinting.
These functions indicate clear intent to fingerprint the user's browser and uniquely identify them.

Next, to fully automate the detection of browser fingerprinting, we modify our puppeteer-based crawler to start with the built-in profiler tool of the Chromium browser, enabled.
This was achieved using Puppeteer's ability to create a session for the Chrome DevTools protocol~\cite{chromeDevTools}.
Additionally, we set the sampling interval of the profiler to 500 $\mu$s, which results to 2K samples per second.
The output of the Chromium profiler is a list of profile nodes.
Each node contains information about samples, in addition to a unique ID and a call frame.
Using this call frame, we extract the function name along with the URL of the JavaScript script that contains the specific function.
This enables us to search for fingerprinting functions, as well as identify the exact script that performs browser fingerprinting.

\point{Limitation:} Although our mechanism is fully automated, we must acknowledge that our fingerprinting detection process may miss minified or obfuscated fingerprinting scripts.

%% file: sections/04_measurements.tex
\begin{figure}[t]
    \centering
    \includegraphics[width=.7\columnwidth]{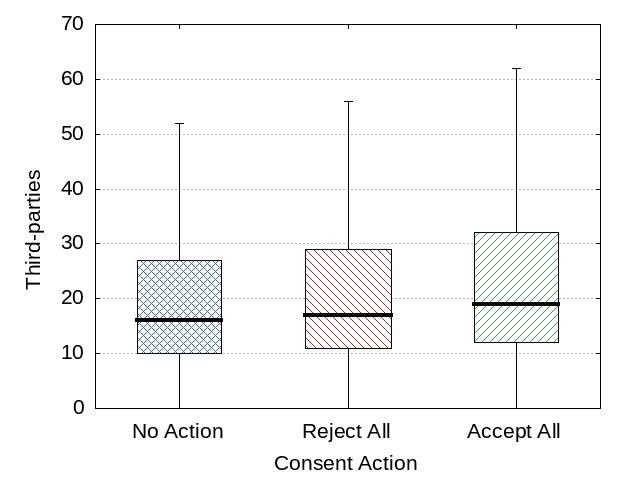}\vspace{-0.3cm}
    \caption{Number of third-parties running on the website during the three different types of visits (\percentileLabels). Surprisingly, for the median website, in the \rejectall case there were more (\ie \thirdpartiesrejectallMedian) third-parties running than in the \noaction case (\ie \thirdpartiesnoactionMedian).}
    \label{fig:ThirdPartiesEcosystem}
\end{figure} 

\section{Analysis of Consent}
\label{sec:measurements}
In this section, we present our measurements and analyze the behavior of websites across three types of visits: when consent is (i) rejected (\rejectall), (ii) granted (\acceptall), and (iii) not responded to (\noaction).

\subsection{Cookie Consent and Third-Parties}
\label{subsec:general-cookie}

Fist, we study how websites change their user tracking behavior depending on the consent provided (or not), via number of third-parties they interact with.
Therefore, we measure the number of unique third-parties running a script on the websites of our dataset before and after a user consent action.
In Figure~\ref{fig:ThirdPartiesEcosystem}, we plot the results (\percentileLabels) for the three user actions.
Two-sided non-parametric Kolmogorov–Smirnov tests for the three cases demonstrated that the three distributions are statistically different, with p-value<$10^{-10}$ ( $D_{\footnotesize{(noaction,rejectall)}}$=$0.038$,
$D_{\footnotesize{(rejectall,acceptall)}}$=$0.061$,
$D_{\footnotesize{(acceptall,noaction)}}$=$0.097$).
As we can see in Fig.~\ref{fig:ThirdPartiesEcosystem}, in the \noaction and in the \acceptall case, there are \thirdpartiesnoactionMedian and \thirdpartiesacceptallMedian third-parties running in the median website, respectively.
Surprisingly, in the \rejectall case, there were more (\ie~\thirdpartiesrejectallMedian) third-parties running in the median website and may reach up to 29 for the 75th percentile.
This suggests that interacting with the consent manager has an impact on the number of third-parties in the visited website.
Specifically, there are more third-parties running in the median website when consent was denied.

\subsection{Sharing User IDs with Third-Parties}
\point{\Idleaking}
In our next experiment, we set out to explore the cases where a first-party ID (\eg cookie, device ID~\cite{nikiforakis2013cookieless}), previously set by the visited website, is getting leaked to third-parties.
Therefore, we measure how many \idleaking operations are being performed in a website as a function of the three aforementioned user choices.
One would expect that there are \emph{no} such operations before the user makes a choice (\ie \noaction), as well as when the users rejects all cookies (\ie \rejectall\/).
However, as shown in Table~\ref{tbl:detected_syncs}, among the websites that present their users with a cookie consent banner, we found 14,238 of them to perform \idleaking even before their users had the opportunity to register their preferences (\noaction case).
To our surprise, when users \rejectall cookies, the \idleaking only gets worse, with more than 15,334 of them engaging in it.

\begin{table}[t]
    \caption{Number of websites detected (i) leaking their first-party user IDs and (ii) having  third-parties that perform synchronizations of user IDs.
    }
    \label{tbl:detected_syncs}
    \footnotesize
    \centering\vspace{-0.3cm}
    \begin{tabular}{lcc}
        \toprule
        {\bf Consent} & {\bf Websites engaging in } & {\bf Websites with third-party} \\
        {\bf Action} & {\bf \idleaking} & {\bf ID synchronization} \\
        \noaction & 14,238 (52.38\%) &  6,533  (24.03\%) \\
        \rejectall & 15,334   (56.41\%) & 7,123  (26.20\%) \\
        \acceptall & 17,764   (65.35\%) &  8,048  (29.61\%) \\ \bottomrule
    \end{tabular}\vspace{-0.2cm}
\end{table}

\begin{table}[t]
    \caption{Average number of unique third-parties learning a user ID. A user's browser leaks first-party IDs to \idleakingnoactionAverage third-parties and  engages on \csyncnoactionAverage synchronizations per third-party ID, on average, even before the user accepted or rejected consent.}
    \label{tbl:averageUniqueThirdParties}
    \centering
    \footnotesize\vspace{-0.3cm}
    \begin{tabular}{lccc}
    \toprule
        \bf ID & \textbf{\noaction} & \textbf{\rejectall} & \textbf{\acceptall} \\ 
    \midrule
        First-party ID & \idleakingnoactionAverage & \idleakingrejectallAverage & \idleakingacceptallAverage \\
        Third-party ID & \csyncnoactionAverage & \csyncrejectallAverage & \csyncacceptallAverage \\
    \bottomrule
    \end{tabular}\vspace{-0.2cm}
\end{table}

\begin{table}[t]
    \caption{Top-5 third-parties that learn the highest number of first-party IDs per consent action in our dataset.}
    \label{tbl:topIdLeakingThirdParties}
    \centering
    \footnotesize\vspace{-0.3cm}
    \begin{tabular}{llll}
    \toprule
    \bf \#  &  \textbf{\noaction} & \textbf{\rejectall} & \textbf{\acceptall} \\ 
    \midrule
     1. & facebook.com & facebook.com & facebook.com \\
        & 18.87\% & 18.29\% & 19.48\% \\
     2. & google-analytics.com & google-analytics.com & google-analytics.com \\
        & 18.85\% & 17.28\% & 15.99\% \\
     3. & bing.com & bing.com & bing.com \\
        & 9.64\% & 8.84\% & 10.27\% \\
     4. & hubspot.com & doubleclick.net & doubleclick.net \\
        & 6.66\% & 6.60\% & 6.82\% \\
     5. & doubleclick.net & hubspot.com & hubspot.com \\
        & 4.68\% & 5.86\% & 5.99\% \\
    \bottomrule
    \end{tabular}\vspace{-0.2cm}
\end{table}

Next, we explore what is the extent of these leaks.
Table~\ref{tbl:averageUniqueThirdParties} shows the average number of \idleaking, as a function of the three user choices.
There are \idleakingnoactionAverage \idleaks even before the user has the opportunity to accept cookies or not (blue bar-\noaction).
To make matters worse, if the user chooses to reject all cookies (red bar-\rejectall), a first-party ID may be leaked to even more third-parties, on average (\idleakingrejectallAverage).
Furthermore, in Table~\ref{tbl:averageUniqueThirdParties}, we measure the average number of third-parties that learn a first-party in the websites we detected this phenomenon.
The difference between \rejectall and \acceptall is rather small: in the average website, choosing \acceptall leaks first-party IDs to \idleakingacceptallAverage third-parties, when \rejectall leaks IDs to \idleakingrejectallAverage third-parties, \ie~about 25\% less.
The difference between the two is hardly significant, implying that more than 75\% of the third-parties that will learn a first-party ID, do so without user's consent!

In Table~\ref{tbl:topIdLeakingThirdParties}, we show the top-5 third-parties in our dataset that learn the most first-party IDs across all websites for each of the three consent options.
Facebook with its social plugin, Google with its analytics tracker and ad-exchange (Doubleclick) modules, and Microsoft (Bing) occupy the top positions in all three consent options.

\begin{quote}
\noindent{{\bf Finding}: Browsers leak more information when users choose to reject all cookies than to take no action at all.
In fact, more than 75\% of the leaks happen despite the fact that users have chosen to reject all cookies.
}
\end{quote}

\begin{table}[t]
    \caption{Top-5 third-parties with highest number of third-party synchronisations per consent action in our dataset.}
    \label{tbl:topCookieSyncThirdParties}
    \centering
    \footnotesize\vspace{-0.3cm}
    \begin{tabular}{llll}
    \toprule
        \bf \# & \textbf{\noaction} & \textbf{\rejectall} & \textbf{\acceptall} \\ 
    \midrule
        1. & doubleclick.net & doubleclick.net & doubleclick.net \\
           & 21.15\% & 21.47\% & 20.22\% \\
        2. & everesttech.net & everesttech.net & everesttech.net \\
           & 13.21\% & 12.10\% & 10.89\% \\
        3. & scorecardresearch.com & facebook.com & facebook.com \\
           & 10.59\% & 9.95\% & 9.61\% \\
        4. & facebook.com & scorecardresearch.com & ad.gt \\
           & 10.15\% & 9.61\% & 9.54\% \\
        5. & taboola.com & google-analytics.com & taboola.com \\
           & 9.68\% & 8.30\% & 8.49\% \\
    \bottomrule
    \end{tabular}\vspace{-0.2cm}
\end{table}

\begin{figure*}[t]
     \centering
        \begin{subfigure}[t]{0.32\textwidth}
        \includegraphics[width=1.1\linewidth]{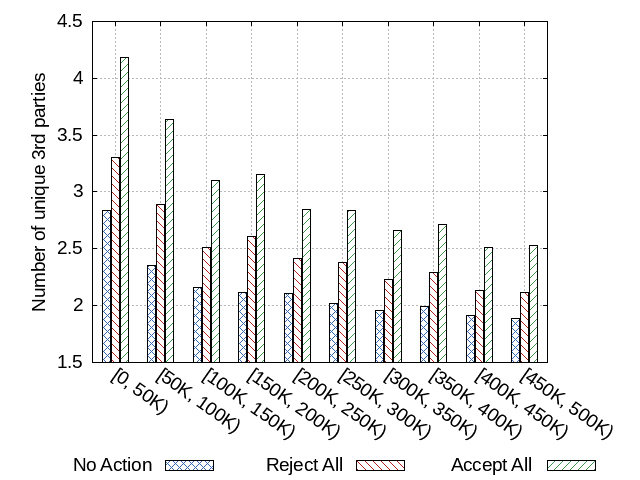}
        \caption{Unique third-parties in ID leaking per website rank.}
        \label{fig:rank}
    	\end{subfigure}
    	\hfill
    	\begin{subfigure}[t]{0.32\textwidth}
        	\includegraphics[width=1.1\linewidth]{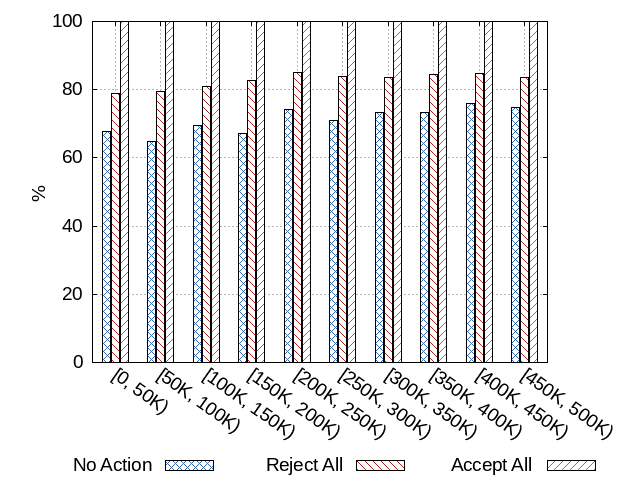}
            \caption{Unique third-parties in ID leaking per website rank (normalized).}
            \label{fig:rank_perc}
        \end{subfigure}    
         \hfill   
        \begin{subfigure}[t]{0.32\textwidth}
            \includegraphics[width=1.1\linewidth]{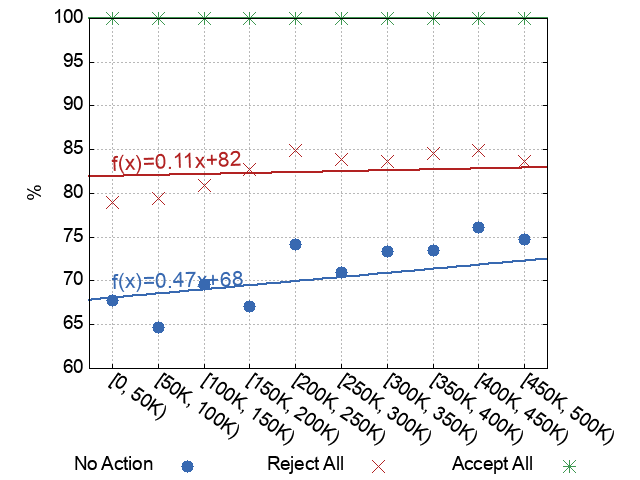}
            \caption{Unique third-parties in ID leaking per website rank (linear fit).}
            \label{fig:rank_trend}
        \end{subfigure}
    \caption{\Idleaking as a function of the website's popularity.
    (a) Average number of unique third-parties involved in ID leaking per rank range of the website
    (b) This figure plots the same information as Fig.~\ref{fig:rank}, with the exception that all \acceptall values are normalized to 100\%.
    (c) This figure plots the same information as Fig.~\ref{fig:rank_perc}, with the exception that all \acceptall values are normalized to 100\%. \rejectall and \noaction points have been fitted with a straight line. The line suggests an increasing trend implying less popular sites are more aggressive at disregarding user choices.}
    \vspace{-0.3cm}
    \label{fig:rank_all}
\end{figure*}

\point{Third-party ID synchronization}
Apart from sharing the first-party IDs they assign to the visiting users, websites also host third-parties that (as described in Section~\ref{sec:csync}) need to synchronize the different user IDs they use for the same users, in order to merge their databases on the back-end.
This way, they can later target specific groups of users~\cite{agarwal2020stop}, sell their data~\cite{selldata}, or use these data in ad-auctions~\cite{10.1145/3131365.3131397,pachilakis2019measuring}.
This type of leakage is worse than the first-party ID leaking, since (1) it is not in the control of the websites themselves, (2) via this mechanism, third-parties \emph{that are not present on the website} can be alerted of a user's presence.

As shown in Table~\ref{tbl:detected_syncs}, from the websites that present their users with a consent manager, we found 6,533 websites hosting third-parties that conduct synchronization of IDs before users had the opportunity to register their choices (\noaction).
If users \rejectall cookies, then even more websites (7,123) engage in ID synchronization.
Although consistent with the finding of the previous subsection (\idleaking), this fact sadly shows that websites employ sophisticated forms of tracking totally disregarding user consent preferences.

To quantify the extent of the phenomenon that happens as a function of the three consent choices examined, in Table~\ref{tbl:averageUniqueThirdParties} we measure the average number of unique third-parties synchronizing a user ID.
When the user takes \noaction, their browser engages in \csyncnoactionAverage synchronizations, on average.
This means that when the user is asked for GDPR compliance, and before even responding, their browser already leaked at least one third-party ID to more than three other third-parties.
To make matters worse, if the user responds negatively and chooses to \rejectall cookies, their cookies may get synced with even more third-parties: \csyncrejectallAverage, on average.

In Table~\ref{tbl:topCookieSyncThirdParties}, we show the top-5 third-parties conducting the highest number of synchronizations across websites, for each of the consent options.
This time, Google's ad-exchange platform {\tt doubleclick.net} and Amazon tracker {\tt everestTech.net} are the top-2 in all three consent options.

\begin{quote}
\noindent{{\bf Finding}: Websites with embedded third-parties that synchronize the IDs they have assigned for the same user, force browsers to engage in \csyncnoactionAverage synchronizations, on average, even before the users had any chance to accept or reject consent.}
\end{quote}

\begin{table}[t]
    \caption{Websites performing browser fingerprinting.}
    \label{tbl:fingerprinting}
    \centering
    \footnotesize\vspace{-0.3cm}
    \begin{tabular}{lrr}
    \toprule
        \textbf{Description} & \textbf{Volume} & \textbf{\% of total} \\ 
    \midrule
        \noaction & 279 & 1.03\% \\
        \rejectall & 285 & 1.05\% \\
        \acceptall & 330 & 1.21\% \\
        In at least one consent action & 336 & 1.24\% \\
    \midrule
        In all 3 cases & 247 & 73.5\% \\ 
        Only in \acceptall case & 47 & 13.9\% \\ 
        Only in \rejectall case & 3 & 0.9\% \\ 
        Wait for action & 7 & 2\% \\ 
        \bottomrule
    \end{tabular}\vspace{-0.2cm}
\end{table}

\subsection{Browser Fingerprinting}
In our next experiment, we set out to explore whether websites track users differently via browser fingerprinting, given the different user responses to the requested cookie consent.
By using the methodology presented in Section~\ref{sec:canvas_det}, we detect the number of websites performing browser fingerprinting across the different types of visit.
Table~\ref{tbl:fingerprinting} presents our findings and, as we can see, the action of the user has no significant impact on the websites' fingerprinting operations.
Specifically, 279 websites perform browser fingerprinting even before the user had the opportunity to respond to the consent request (\ie \noaction).
Even worse, if the user chooses to \rejectall cookies, even more websites engaged in browser fingerprinting: 285 websites.
Interestingly, we see 73.5\% of the fingerprinting websites perform browser fingerprinting no matter what the user consent action is.
In addition, we see that only 2\% of these websites wait for user's action before starting their fingerprinting operation.
Only 13.9\% of them perform browser fingerprinting only when the user gives consent, and 0.9\% of the websites perform browser fingerprinting \emph{only if the user rejects giving consent}.
It is apparent that these websites are using browser fingerprinting as a fallback mechanism in case they are not allowed (by the GDPR) to set a cookie on the user side.
It is important to stress at this point that based on Article 4/Recital 30~\cite{recital30}, GDPR regards the process of user identifying information and not cookies per se.

\begin{figure*}[t]
     \centering
        \begin{subfigure}[t]{0.32\textwidth}
            \includegraphics[width=1.1\columnwidth]{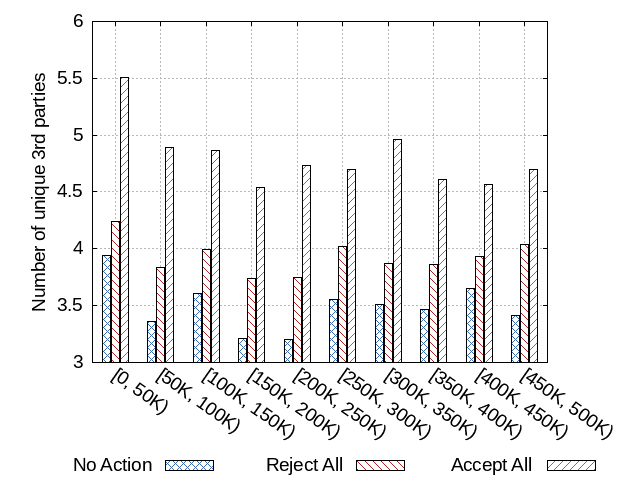}
            \caption{Unique third-parties involved in ID synchronizations per website rank.}
            \label{fig:cs_rank}
        \end{subfigure}
        \hfill
        \begin{subfigure}[t]{0.32\textwidth}
            \includegraphics[width=1.1\columnwidth]{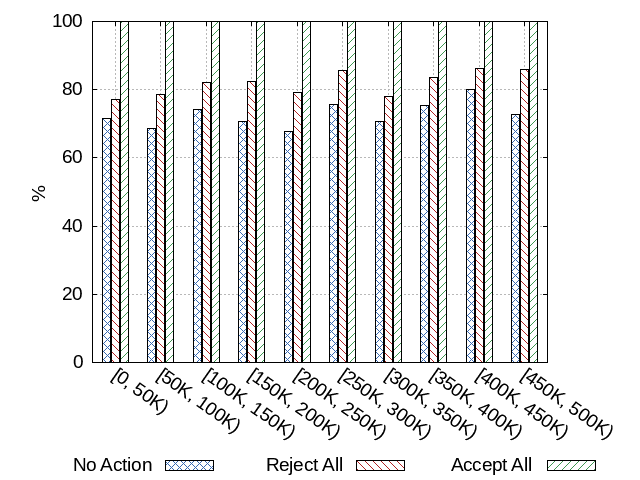}
            \caption{Unique third-parties involved in ID synchronizations per website rank (normalized).}
            \label{fig:cs_rank_perc}
        \end{subfigure}
        \hfill
        \begin{subfigure}[t]{0.32\textwidth}
            \includegraphics[width=1.1\columnwidth]{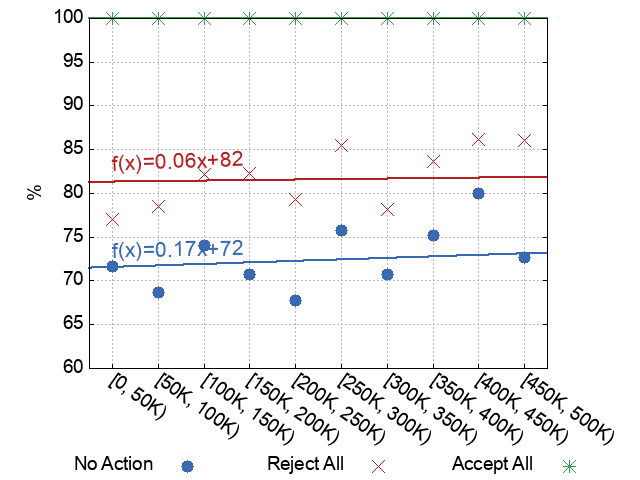}
            \caption{Unique third-parties involved in ID synchronizations per website rank (linear fit).}
            \label{fig:cs_rank_trend}
        \end{subfigure}
    \caption{ID Synchronization as a function of the website's popularity.
    (a) Average number of unique third-parties involved in synchronizations per rank range of the website.
    (b) This figure plots the same information as Fig.~\ref{fig:cs_rank}, with the exception that all \acceptall values are normalized to 100\%.
    (c) This figure plots the same information as Fig.~\ref{fig:cs_rank_perc}, with the exception that all \acceptall values are normalized to 100\%.
    \rejectall and \noaction points have been fitted with a straight line. We see that the line suggests an increasing trend implying that less popular sites are more aggressive at disregarding user choices.}
    \label{fig:cs_rank_all}
    \vspace{-0.3cm}
\end{figure*}

\begin{quote}
\noindent{{\bf Finding}: Although websites ask users for cookie consent, they do not take into account this consent when they perform browser fingerprinting.}
\end{quote}

\begin{figure*}[t]
     \centering
        \subfloat[Number of unique third-parties learning a first-party ID.]
        {
            \hspace{0.4cm}
            \includegraphics[width=0.85\linewidth]{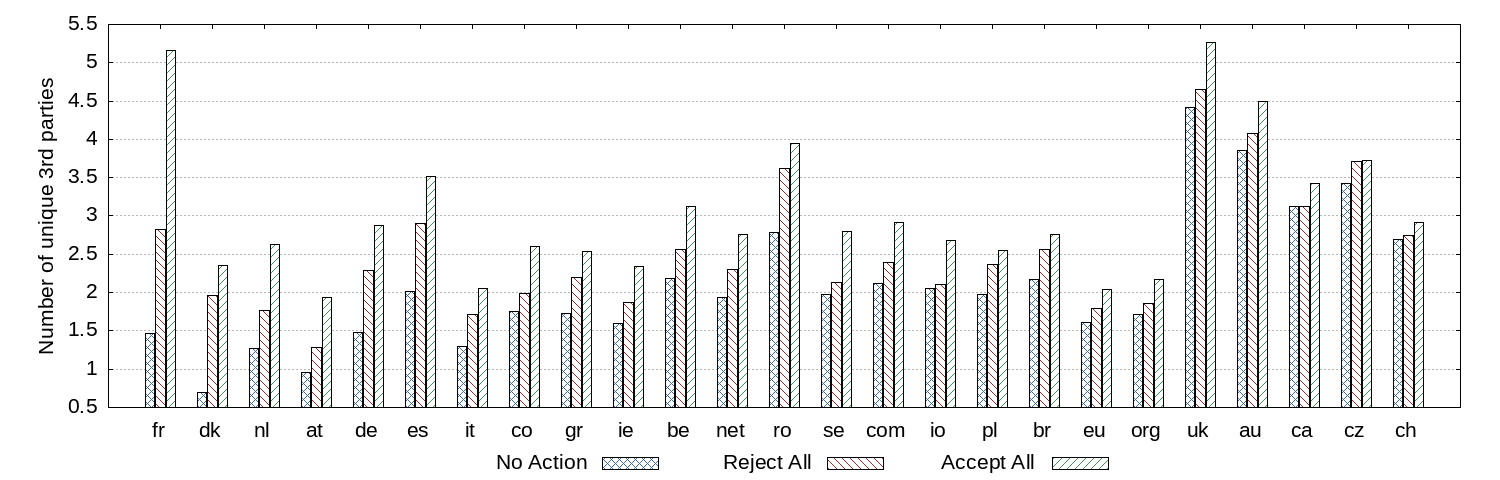} 
            \label{fig:tld_norm}
        }
        \newline
        \subfloat[Normalized number of unique third-parties learning a first-party ID.]
        {
            \includegraphics[width=0.85\linewidth]{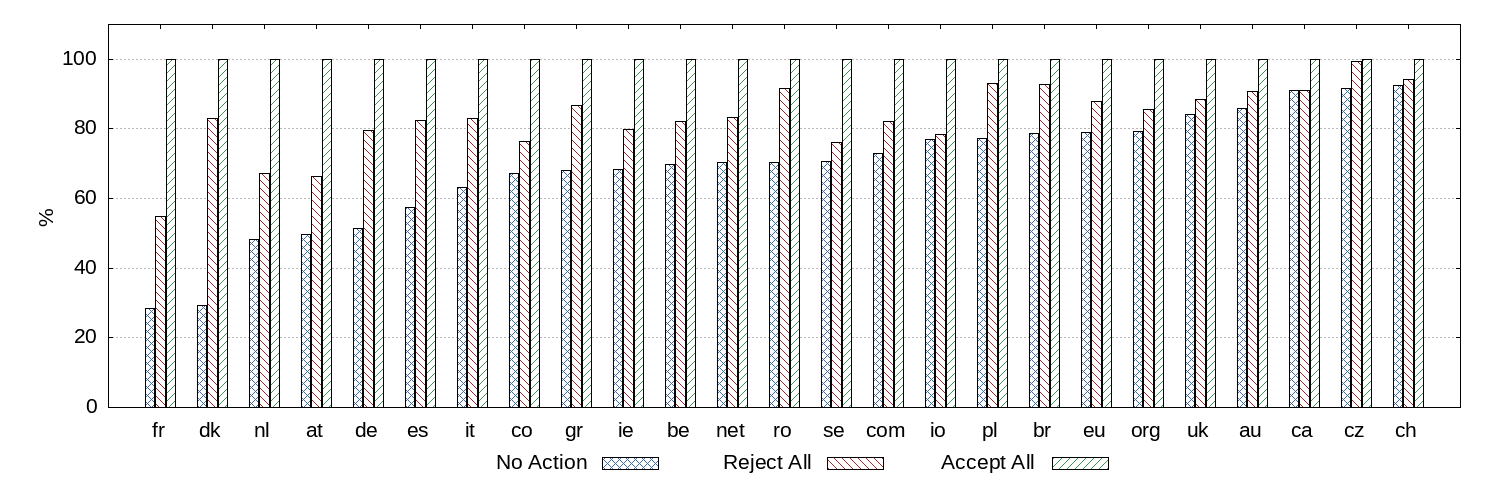} \label{fig:tld_perc}
        }
    \label{fig:tld}
    \caption{Number of unique third-parties learning a first-party ID as a function of the top-level domain per country code.
    (a) This figure plots the absolute values.
    (b) This figure plots the same information as in (a), with the difference that the max value (\acceptall) is normalized to 100\%.
    This enables us to compare websites that have different magnitudes of leakage.
    We see that websites in different domain names have very different behavior.
    For example websites in \emph{.fr} make 1.5 leaks before the user gives consent, close to 2.8 leaks when the user rejects all cookies, and more than 5 leaks when the user accepts all cookies.
    On the other end of the spectrum, user choices in websites in \emph{.cz} seem to have little impact: they leak to 3.7 third-parties both in cases when users choose to \rejectall cookies, and in cases where users choose to \acceptall cookies.}
    \vspace{-0.3cm}
\end{figure*}

\begin{figure*}[t]
        \subfloat[Number of unique third-parties engaged in third-party ID synchronization.]
        {
            \hspace{0.4cm}\includegraphics[width=0.85\linewidth]{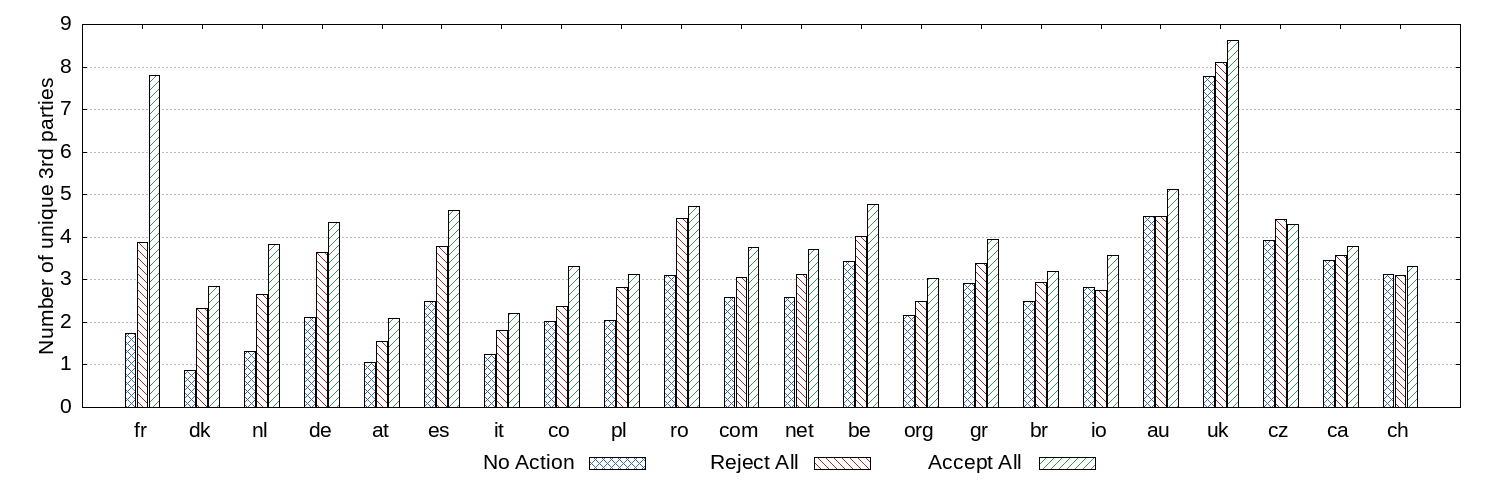} 
            \label{fig:tld_cs}
        }
        \newline
        \subfloat[Normalized number of unique third-parties engaged in ID synchronization.]
        {\centering
            \includegraphics[width=0.85\linewidth]{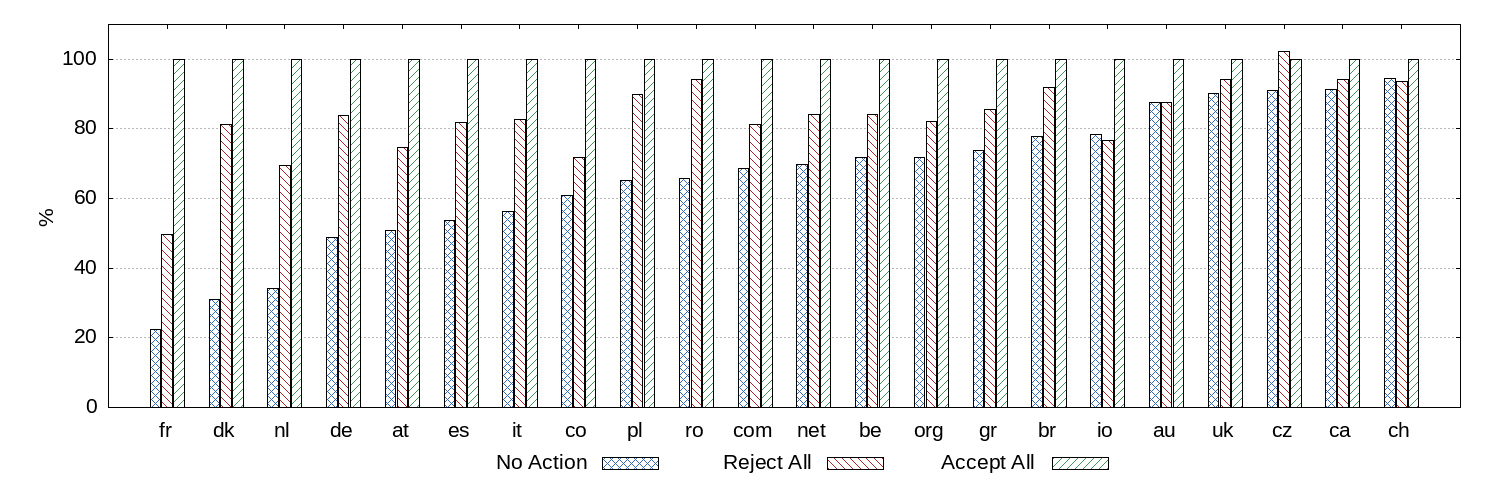} 
            \label{fig:tld_cs_perc}
        }
    \label{fig:tld_cs_main}\vspace{-0.3cm}
    \caption{Number of unique third-parties engaged in third-party ID synchronization as a function of the top-level domain per country code.
    (a) This figure plots the absolute values.
    (b) This figure plots the same information as in (a), with the difference that the max value (\acceptall) is normalized to 100\%.
    This enables us to compare sites that have different magnitudes of leakage.
    As in \idleaking, websites in \emph{.fr} engage in less third-party ID synchronization without the user's consent.
    On the other end of the spectrum, user choices in sites in \emph{.cz} seem to have little impact: they engage in approximately 4.4 third-party ID synchronizations both when users choose to \rejectall cookies, and in cases where users choose to \acceptall cookies.}
\end{figure*}

\subsection{Does website popularity matter?}
In our next experiment, we explore whether a website's popularity impacts how the website respects the user's choices.
For this reason, we grouped the websites into buckets based on their popularity: the first bucket contains the top 50K websites in the Tranco list, the next bucket contains the next 50K most popular sites (\ie~ranking 50K-100K), \etc
In Figure~\ref{fig:rank}, we show the extent of \idleaking for the different buckets for the three cases we study: \noaction (blue bar), \rejectall (red bar), and \acceptall (green bar).
We see that as the popularity of the website decreases (right part of the plot), all bars tend to decrease, implying that the magnitude of tracking through \idleaking decreases as well.
In Figure~\ref{fig:rank_perc}, we normalize the values (so that the \acceptall corresponds to the same 100\% value).
We see that in this case, blue bars and red bars tend to have a slightly increasing trend to the right.
That is, less popular websites tend to be slightly more aggressive in disregarding user choices.
For example, popular websites (0, 50K) do 67\% of their \idleaking before the user makes any choice, while less popular sites (400K, 450K) make 77\% of their \idleaking before the user makes any choice. 

In Figure~\ref{fig:rank_trend}, we show an interpolation of the results using a straight line.
In both consent actions (\noaction and \rejectall), we see a positive slope ($R^2$=$0.42$ and $R^2$=$0.04$, respectively).
Similarly, in Figure \ref{fig:cs_rank_all}, we see the same trend across the popularity buckets of websites hosting synchronizing third-parties: less popular websites (towards the right-end of the figures) tend to be more aggressive at disregarding user choices.

\begin{quote}
\noindent{{\bf Finding}: Less popular websites are more aggressive at disregarding users' consent choices and engage in \idleaking and third-party ID synchronizations.}
\end{quote}

\subsection{Does the hosting country matter?}
Next, we study how the websites hosted in different countries (represented by their country code top-level domain (ccTLD)) treat the user consent.
In Figure~\ref{fig:tld_norm}, we present the results for the case of first-party ID leaks.
As we see, a higher percent of Europe-based ccTLDs respect the choices of the users (\ie less \idleaking): websites with ccTLD=\emph{fr} (France), \emph{dk} (Denmark), \emph{nl} (Netherlands), \emph{at} (Austria) and \emph{de} (Germany), leak first-parties to less number of third-parties than websites with non Europe-based ccTLDs (\eg~\emph{uk} (UK), \emph{ca} (Canada), \emph{au} (Australia)), where the choices of the user do not have any impact.

In Figure~\ref{fig:tld_perc}, we normalize the results based on the \acceptall.
Websites on the right part of the figure tend to disrespect users choices: the difference between \acceptall and \rejectall in sites ending in \emph{.cz} and \emph{.ch} seems to be negligible.
Thus, whether the user chooses \acceptall or \rejectall makes little difference.
On the contrary, websites on the left part of the figure seem to respect user choices more.
For example, the difference between \acceptall and \rejectall for \emph{.fr} websites seem to be close to 50\%.
Similarly, the difference between \noaction and \acceptall for \emph{.fr} websites seems to be more than 70\%.
Surprisingly, we see the ccTLD of \emph{.eu} being on the right side of the figure, which means that there is an increased number of websites in this ccTLD, not yet compliant with GDPR.
Thus, although not perfect, user choices for the websites on the left part of the figure have a meaningful effect, in contrast to websites on the right part.

Similarly, in Figure \ref{fig:tld_cs_perc}, we plot the same results for third-party ID synchronization.
We see that, again, European ccTLDs: \emph{.fr}, \emph{.dk}, \emph{.nl}, \emph{.at} and \emph{.de}, tend to perform less third-party ID synchronization when there is no consent given by the user, than websites with non Europe-based ccTLDs: (\eg~\emph{uk} (UK), \emph{ca} (Canada), \emph{au} (Australia)), where the user choices have a much smaller impact.
Surprisingly, we see two European ccTLDs, \emph{.gr} (Greece) and \emph{.cz} (Czech Republic), not performing like other European ccTLDs.
To make matters worse, websites of \emph{.cz} perform more synchronizations when users deny giving consent.

%% file: sections/05_edge_observations.tex
\section{Ineffective Consent: Edge cases}
\label{subsec:edge-cases}

In our dataset, we observed 73 websites that interact with over 100 unique third-parties each, in at least one of the three types of visit.
One such example with extreme behavior is \texttt{laprovence.com}.
When a user visits the website and gives \acceptall consent, the website interacts with 159 different third-parties and performs synchronization for multiple IDs with 59 of these parties.
We observed the values of 37 unique third-party cookies being leaked to third-parties different from the cookie's owner.
In the \rejectall case, the website interacts with 80 third-parties, and performs synchronization for at least one ID with 16 for them.
Interestingly, when the user lands on the website with a clean session and performs \noaction, but simply waits, the website interacts with 97 third-parties and performs synchronization with 29 of them.

Regarding first-party ID leaking, we observed that multiple websites store a cookie labeled as ``necessary'', but then proceed to leak its value to various third-parties.
For example, \texttt{harryanddavid.com} leaks the values of 28 different first-party cookies in the \rejectall and \noaction cases.
Also, \texttt{diariodepontevedra.es} and \texttt{asivaespana.com}, in the \rejectall and \acceptall cases, respectively, perform ID leaking with 38 different third-parties for more than one ID.

In addition, \texttt{camer.be} interacts with 91 unique third-parties in the \noaction case, 94 in the \acceptall case, and surprisingly with 131 in the \rejectall case.
For the \rejectall case, this website is also involved in a major third-party ID synchronization operation.
At the time of crawling, the website interacted with the third-party \texttt{taboola.com}, which stored a cookie with name \texttt{t\_gid} and value \texttt{884d05cc-335c-4226-ab94-7ab6114fef6a-\linebreak~tuct65bfbc8}.
This value was sent to 20 other third-parties.
One interesting finding is that this cookie is stored only when the user declines consent (\ie \rejectall).

Similarly, \texttt{cnnturk.com} is also involved in a major third-party ID synchronization operation.
Specifically, when the user lands on the website, a third-party called \texttt{lijit.com} stores the cookie \texttt{\_ljtrtb\_42}.
The value of this cookie is then sent to 21 other third-parties.
Interestingly, this behavior is observed only after the user has interacted with the cookie consent form (\ie \rejectall and \acceptall cases).
One example value of this cookie that we observed during the \acceptall case is \texttt{c98d9202-8774-4e11-8c90-99d9cb879930-\linebreak~tuct65c0de5}, which can be used to uniquely identify a user.
Note that \texttt{lijit.com} is an ad-serving domain, which can be found in multiple blacklists for tracking domains.

Finally, \texttt{glamour.com} leaks a unique identifier which is set as the value of a first-party cookie.
Specifically, when a user lands on the website, a cookie called \texttt{CN\_xid} is stored, with one example value being \texttt{73a4ff1f-ff45-4943-bdaa-73658b00bd42}.
Then, this value is sent to exactly 21 unique third-parties.
The third-parties that receive the value of the cookie are exactly the same for all 3 types of visits.
An interesting finding is that the third-parties that receive this value are not only domains known for advertising and analytics (\eg~\texttt{google-analytics.com} and \texttt{securepubads.g.doubleclick.net}), but also legitimate and mainstream websites like \texttt{vogue.com} and \texttt{wired.com}.

%% file: sections/06_related.tex
\section{Related Work}
\label{sec:related}

The recent increased interest of regulators and governments around the privacy rights of Internet users did not result only in rules like GDPR and California Consumer Privacy Act (CCPA), but also in an important body of research.
In~\cite{fouad:hal-02567022}, authors investigate the legal compliance of purposes for 20K third-party cookies collected.
Their findings show that purposes declared in cookie policies do not comply with the purpose specification principle in 95\% of cases.
In~\cite{dabrowski2019measuring}, authors collect cookies from the Alexa Top 100K websites and compare their cookie behavior from different vantage points, to investigate whether there are differences in cookie setting when accessing Internet services from different jurisdictions.
Additionally, they study whether cookie setting behavior has changed over time by comparing today's results with a dataset from 2016.

In~\cite{10.1145/3321705.3329806} authors perform an evaluation of the tracking performed in 2K high-traffic websites, hosted both inside and outside the EU.
Specifically, they evaluate the information presented to users and the actual tracking implemented through cookies.
Their results show that the GDPR has impacted website behavior in a truly global way.
US-based websites behave similarly to EU-based ones, while third-party opt-out services reduce the amount of tracking, even for websites which do not put any effort in respecting GDPR.
On the other hand, they show that cookies can identify users when visiting more than 90\% of the websites they crawled, and they encountered a large number of websites that present deceiving information, making it it very difficult, if at all possible, for users to avoid being tracked.
Similar to this work, in~\cite{eijk2019impact}, authors crawl 1.5K EU, US, and Canadian websites from 18 countries and analyze the cookie notices they find.
Using a series of regression models, they find that a website's Top Level Domain explains a substantial portion of the variance in cookie notice metrics, but the users vantage point does not, which means that websites follow one set of privacy rules for all their users.

In~\cite{utz2019informed}, authors study the common properties of the graphical user interface of consent notices and conduct three experiments with more than 80K unique users on a German website, to investigate the influence of notice position, type of choice, and content framing on consent.
Their results show that (i) users are more likely to interact with a notice shown in the lower left part of the screen, (ii) users are willing to accept tracking compared to mechanisms that require them to allow cookie use for each category or company individually, (iii) the wide-spread practice of nudging has a large effect on the choices users make.
In~\cite{urban2018unwanted} authors study the impact of the legislation on cookie syncing between third-parties.
They show that the general structure of how the entities are arranged is not affected by the GDPR, but the new regulation has a statistically significant impact on the number of connections that shrunk by 40\% in the GDPR era.

In an effort closest to ours, Matte \etal analyzed the GDPR and ePrivacy Directive across 23K European websites to identify legal violations in implementations of cookie banners based on the storage of consent~\cite{9152617}.
That is, they (i) capture the user's choice (consent or not), (ii) measure whether the websites register the same response as the user's choice, (iii) measure whether websites register any response \emph{before} the users click their preference.
They found that: 141 websites register positive consent even if the user has not made their choice; 236 websites nudge the users towards accepting consent by pre-selecting options; and 27 websites store a positive consent even if the user has explicitly opted out.
Performing extensive tests on 560 websites, they found at least one violation in 54\% of them.
Although our work and~\cite{9152617} share similar goals, they clearly have significant differences.
First, although~\cite{9152617} focuses on cookies as the main tracking mechanism, in this work, we focus on post-cookie tracking mechanisms including browser fingerprinting, ID leaking, and ID synchronization.
In this aspect, we explore whether sites use such post-cookie tracking mechanisms to bypass any consent the user has provided for cookies.
Second,~\cite{9152617} focuses on whether the Cookie Management Provider registers the same response as the user's input.
We follow a different methodology and measure \emph{not} the response registered, but the actual tracking mechanisms that are activated when the users access a website.

%% file: sections/07_discussion.tex
\section{Discussion}

\point{GDPR compliance}
One question that comes to mind is whether these websites are in violation of the GDPR and the ePrivacy Directive.
Obviously, one cannot make such an umbrella statement for all the websites studied in this paper.
Such violations should be studied on a case-by-case basis.
Even further, each website is different, and may have a legal basis to collect user data that goes beyond the user consent.
What we identify in this paper is a \emph{disparity} between (i) what the users perceive about the collection of their data, and (ii) what some websites implement with respect to data processing.
Indeed, by being shown a cookie consent banner, users perceive that they are being asked to give their permission to the website to collect and process their data.
Even further, when they are given several choices, users feel that they are empowered to give a fine-grain permission, which will obviously be taken into account.

Unfortunately, this perception of the users is completely different from what various websites implement. 
In this paper, we saw that several websites collect (and share with third-parties) information about their users, even before the users had the opportunity to register their preference.
Even worse, when the users said that they would like to reject all cookies, collection of their data even intensified.
Indeed, each website is ultimately responsible for the consent asked from their visitor.
However, it is not obvious if the legal responsibility is shifted to the Consent Management Platform (CMP).
Nonetheless, and considering our results, it is hard to believe that all these publishers do not respect the users' consent choices without intention (\eg~due to software bug, bad developer practices or wrong integration with their CMP).

Interestingly, existing literature, websites and blog-posts around the GDPR and changes it brought on the Internet and user tracking~\cite{solomos2020gdpr-changes}, focus solely on how identifiable information stored in cookies is maintained.
However, as highlighted here, the GDPR is not only about cookies.
Instead, we aim to increase user awareness regarding the GDPR (non)compliance of deployed stateless (\ie~cookie-less) tracking, and influence a change in language used in consent request statements, to be GDPR-compliant and reflect closely what the websites do in reality, in comparison to what is explained to the user.

Furthermore, our analysis of tracking per country code reveals significant discrepancies across EU (or not) countries.
These results highlight the lack of effort from specific local governments regarding the digital privacy rights of their citizens.
Our results can motivate them to take action and increase the GDPR enforcement to make websites hosted in their countries aligned with the rest of EU countries with respect to the GDPR compliance.

\point{Inbound vs. Outbound Information}
Although user tracking without user consent is generally undesirable, in this paper, we studied some sophisticated approaches to user tracking (such as \idleaking and ID synchronization) which involved not only data collection, but also data sharing with third-parties.
Indeed, both approaches, provide to third-parties identifiers associated with the current user.
In this way, third-parties will be able to know that this user has visited the specific website (even if they are not embedded in that website).
And this happens even before the user has given any permission for data collection on the cookie consent banner.
To put it simply: the website has already told third-parties that this user has just visited, while the user still makes up their mind whether to give consent for data collection or not.
Thus, the user is asked for consenting to something that has already happened and it will keep on happening even if the user denies consent.

\point{Edge Cases}
Someone could argue that the edge-cases studied in this paper are momentary, and cannot be held against websites as proof of non-GDPR compliance.
However, even though we acknowledge the dynamicity of websites, we made a best effort to provide results that were repeatable across multiple crawls.
In fact, changes in third-parties embedded in a website could change their intensity of tracking.
We anticipate such changes are transient and infrequent in websites, and that high intensity of tracking is repeatable.

\point{Methodology}
The methodology we presented in this paper can be transformed into an auditing tool for regulators, stakeholders and privacy-policy makers, for verifying compliance with the GDPR, ePrivacy Directive, and users' privacy rights.
Our approach links together the
(i) requested user consent of webmaster with
(ii) actions taken by the website based on the particular consent given.
Apart from these entities, browser vendors have already shown interest in blocking bad policies on websites~\cite{cookiebot2021chrome-cookies,wilander2019tracking-prevention,dunn2020chrome-blocking-downloads} and our methodology can help towards exactly these goals.
Specifically, by following our methodology, browser vendors can detect at run-time stateless device fingerprinting attempts~\cite{brave2020} and compare these actions with given user consent.

%% file: sections/08_conclusion.tex
\section{Conclusion}
\label{sec:conclusion}

Over the past couple of years, an increasing number of websites have started to present users with cookie consent banners: pop-up windows that ask for user's permission to send/receive cookies.
Such banners provide a variety of choices including (i) accept all, (ii) reject all, and (iii) accept some cookies.
In this paper, we study whether these websites that present users with cookie consent banners, track their users using ``non cookie'' approaches including \idleaking, third-party ID synchronization and browser fingerprinting.

In our experiments, we found 15,334 websites that track their users using \idleaking.
Even further, this tracking happened despite the fact that users of these websites had rejected all cookies in the cookie consent banner! 
In fact, most of these websites (14,238) had started the \idleaking tracking even before the users had any opportunity to register their consent choice.

Therefore, we highlight a gap between what users expect to happen when they see a cookie consent banner and what several websites do as a result of users' choices.
We feel that research like this helps increase transparency on the Web and expose websites which do not correspond to users' expectations, and are non-GDPR compliant.

Future work could focus on even harder questions such as:
How should third-parties connect into CMP prompts?
Is it intentional that some third-parties only take action on ``reject all'' option? If yes, why?
Are some CMPs better than others with respect to GDPR compliance?
Are all these privacy violations the website's, the CMP's, or the third-party's fault?

%% file: sections/09_acknowledgments.tex
\begin{acks}
This project received funding from the EU H2020 Research and Innovation programme under grant agreements No 830929 (CyberSec4Europe), No 830927 (Concordia), No 871793 (Accordion), and No 871370 (Pimcity).
These  results reflect only the authors' view and the Commission is not responsible for any use that may be made of the information it contains.
\end{acks}